\documentclass{elsart}
\newcommand{\be}{\begin{equation}}
\newcommand{\ee}{\end{equation}}
\newcommand{\beq}{\begin{eqnarray}}
\newcommand{\eeq}{\end{eqnarray}}
\newcommand{\mbf}{\mathbf}

\begin{document}

\begin{frontmatter}

\title{Analyzing Multiple Nonlinear Time Series with Extended Granger Causality}

\author{Yonghong Chen}

\address{Xi'an Jiaotong University, Xi'an 710049, P.R. China}
\address{Center for Complex Systems and Brain Sciences, Florida Atlantic University, Boca
Raton, FL 33431, USA} \ead{ychen@walt.ccs.fau.edu}

\author{Govindan Rangarajan}

\address{Department of Mathematics and Centre for Theoretical Studies,
Indian Institute of Science, Bangalore 560 012, India}
\address{Jawaharlal Nehru Centre for Advanced Scientific Research,
Bangalore, India} \ead{rangaraj@math.iisc.ernet.in}

\author{Jianfeng Feng}

\address{Department of Informatics, Sussex University, Brighton BN1 9QH, United Kingdom}
\ead{jianfeng@cogs.susx.ac.uk}

\author{Mingzhou Ding}

\address{Center for Complex Systems and Brain Sciences,
Florida Atlantic University, Boca Raton, FL 33431, USA}
\ead{ding@fau.edu}

\newpage

\begin{abstract}
Identifying causal relations among simultaneously acquired signals
is an important problem in multivariate time series analysis. For
linear stochastic systems Granger proposed a simple procedure
called the Granger causality to detect such relations. In this
work we consider nonlinear extensions of Granger's idea and refer
to the result as Extended Granger Causality. A simple approach
implementing the Extended Granger Causality is presented and
applied to multiple chaotic time series and other types of
nonlinear signals. In addition, for situations with three or more
time series we propose a conditional Extended Granger Causality
measure that enables us to determine whether the causal relation
between two signals is direct or mediated by another process.
\end{abstract}

\end{frontmatter}

\newpage
\section{Introduction}

Given the deluge of multi-channel data generated by experiments in
both science and engineering, the role of multivariate time series
analysis, especially nonlinear time series processing, has become
crucial in understanding the patterns of interaction among
different elements of nonlinear systems. In particular,
identifying causal relations among signals is important in fields
ranging from physics to biology to economics. One approach to
evaluating causal relations between two time series is to examine
if the prediction of one series could be improved by incorporating
information of the other. This was originally proposed by Wiener
\cite{wiener} and later formalized by Granger in the context of
linear regression models of stochastic processes \cite{granger}.
Specifically, if the variance of the prediction error of the
second time series at the present time is reduced by inclusion of
past measurements from the first time series in the linear
regression model, then the first time series is said to have a
causal influence on the second time series. The roles of the two
time series can be reversed to address the question of causal
influence in the opposite direction. From this definition it is
clear that the flow of time plays a vital role in making direction
related inference from time series data.

Since Granger causality was formulated for linear models, its
direct application to nonlinear systems may or may not be
appropriate, depending on the specific problem. In some cases, the
linear Granger causality is able to identify the correct patterns
of interaction for multiple nonlinear time series, but in some
other cases, as will be shown later in this paper, it fails to do
so. We deal with this issue by extending Granger's idea to
nonlinear problems. Our starting point is the standard delay
embedding reconstruction of the phase space attractors. Clearly, a
full description of a given attractor requires a nonlinear set of
equations. But, locally, one can approximate the dynamics
linearly. Applying Granger's causality idea to each local
neighborhood and averaging the resulting statistical quantity over
the entire attractor results in Extended Granger Causality Index
(EGCI). We examine the effectiveness of this idea on numerically
generated nonlinear time series with known patterns of
interaction.

Works related to the identification of interdependence in
nonlinear systems have appeared in the literature
\cite{prie,wend,frei,schrei,saka}. Particularly relevant for the
work in this paper are works based on delay coordinate embedding
reconstruction of phase space. Along this direction a number of
methods of detecting nonlinear interdependence or coupling based
on nonlinear prediction theory have appeared in the past few years
\cite{schreiber,arnhold,quian1,quian2,quyen,bhatt,break1,break2,schiff,wies}.
The basic ideas in these papers are similar, all involving the use
of the points in a neighborhood in the reconstructed space to
predict future dynamical behavior. In
\cite{arnhold,quian1,quian2,quyen,bhatt,break1,break2,schiff},
time indices of neighborhood points in the space $\mbf{X}$
reconstructed from one time series ($x$) are used to predict the
dynamics in space $\mbf{Y}$ reconstructed from the second time
series $y$. If this prediction is good enough, then it implies a
dependence from $x$ to $y$. Similarly, the reverse dependence can
be found. Different authors define different criteria to quantify
the goodness of the prediction, but the common assumption that
nearby points in one reconstructed space $\mbf{X}$ map to nearby
points in another reconstructed space $\mbf Y$ is adopted. These
methods tend to detect strong interactions such as
synchronization, phase synchronization or generalized
synchronization. In order to detect weak interactions, a
modification\cite{wies} was made by presenting a mixed-state
prediction method where a reconstruction of mixing two time series
was employed. It is important to note that all these nonlinear
prediction based methods employ the same kind of predictor (a
zeroth order predictor) which takes the mean or weighted mean as
the prediction value. Since points in a given neighborhood come
both from the past and the future of the reference point this kind
of prediction does not account properly  for the flow of time. Our
idea differs from the previous methods in two main respects: (a)
an linear regression predictor is employed for each local
neighborhood and (b) as a consequence the flow of time is
explicitly incorporated in the predictor which is an essential
element of inferring causal relations in multiple time series
\cite{granger}. A nonlinear approach that shares a number of
similarities with ours has appeared in \cite{frei}.

\section{Theory}

In this section we will first review the basic idea of Granger
Causality formulated for analyzing linear systems and then propose
a generalization of Granger's idea to attractors reconstructed
with delay coordinates.

{\it Granger Causality}: The method of detecting causal relations
among multiple linear time series is based on linear prediction
theory. For a stationary time series $x(t)$, consider the
following AutoRegressive (AR) prediction of the current value of
$x(t)$ based on $m$ past measurements: \be \label{uniprd}
x(t)=\sum^{m}_{j=1} \alpha_j x(t-j)+\varepsilon_x(t).\ee Here
$\varepsilon_x(t)$ is the prediction error whose magnitude can be
evaluated by its variance $var(\varepsilon_x(t))$. Suppose that
simultaneously we have also acquired another stationary time
series $y(t)$. Consider the following prediction of the current
value of $x(t)$ based both on its own past values and the past
values of $y(t)$: \be \label{biprd} x(t)=\sum^{m}_{j=1} a_j
x(t-j)+\sum^{m}_{j=1} b_{j} y(t-j)+\varepsilon_{x|y}(t).\ee If the
prediction improves by incorporating the past values of $y(t)$,
that is, $var(\varepsilon_{x|y}(t))<var(\varepsilon_{x}(t))$ in
some suitable sense, then we say that $y(t)$ has a causal
influence on $x(t)$. Similarly, we may consider \beq
y(t) & = & \sum^{m}_{j=1} \beta_j y(t-j)+\varepsilon_y(t),\\
y(t) & = & \sum^{m}_{j=1} c_j x(t-j)+\sum^{m}_{j=1} d_{j}
y(t-j)+\varepsilon_{y|x}(t) \label{biprdy}. \eeq and say that
$x(t)$ has a causal influence on $y(t)$ if
var($\varepsilon_{y|x}(t))<var(\varepsilon_{y}(t))$. We note that
Eqs. (\ref{biprd}) and (\ref{biprdy}) together form the following
Vector AutoRegressive model (VAR):

\parbox{10cm}{ \beq
\label{granger} x(t)=\sum^{m}_{j=1}
a_j x(t-j)+\sum^{m}_{j=1} b_{j} y(t-j)+\varepsilon_{x|y}(t),\nonumber \\
y(t)=\sum^{m}_{j=1} c_j x(t-j)+\sum^{m}_{j=1} d_{j}
y(t-j)+\varepsilon_{y|x}(t), \nonumber \eeq} \hfill
\parbox{1cm}{\beq  \eeq}

where standard techniques exist to estimate such models from time
series data.

{\it Extended Granger Causality:} Consider two nonlinear time
series $x(t)$ and $y(t)$. The joint dynamics is reconstructed with
the following delay vector \cite{casdag,pecora} \be \label{vector}
\mbf{z}(t)=(\mbf x(t)^T, \mbf y(t)^T)^T,\ee where $\mbf
x(t)=(x(t), x(t-\tau_1),\cdots,x(t-(m_1-1)\tau_1))^T, \ \ \mbf
y(t)=(y(t),y(t-\tau_2), \cdots,y(t-(m_2-1)\tau_2))^T,$ $m_i$ is
embedding dimension and $\tau_i$ is time delay for $i=1,2$.
Usually, the embedding dimensions and time delays for different
series can be different.  However when we investigate Granger
causality, the time delays must be equal so that causal inferences
can be made.  Hereafter we take $\tau_1=\tau_2=\tau$.

In the delay embedding space, there exists a function $\mbf f$
that maps a given point $\mbf z(t)$ to its observed image $\mbf
z(t+\tau)$. Usually, this function has no known analytical form
but can be locally approximated by a linear map around some
reference point \cite{eckmann,farmer}: $\mbf z(t+\tau)=\mbf A \mbf
z(t)+ \mbf{r}(t)$, where $\mbf A$ is $(m_1+m_2)\times(m_1+m_2)$
coefficient matrix which can be determined by the least-squares
technique and $\mbf r(t)$ is the error vector. Substituting
Eq.(\ref{vector}) in the above relations, we get the following
equations:
\beq \label{linear} \left(%
\begin{array}{c}
  x(t+\tau) \\
  y(t+\tau) \\
\end{array}%
\right) & = & \mbf A_1  \left(%
\begin{array}{c}
  x(t) \\
  y(t) \\
\end{array}%
\right) + \mbf A_2  \left(%
\begin{array}{c}
  x(t-\tau) \\
  y(t-\tau) \\
\end{array}%
\right) + \nonumber \\
& \ \ & \cdots +  \mbf A_{m}  \left(%
\begin{array}{c}
  x(t-(m-1)\tau) \\
  y(t-(m-1)\tau) \\
\end{array}%
\right) + \left(%
\begin{array}{c}
  \varepsilon_{x|y} \\
  \varepsilon_{y|x} \\
\end{array}%
\right) ,
\eeq
where $\mbf A_i=\left(%
\begin{array}{cc}
  a_{11}^{(i)} \ \ & a_{12}^{(i)} \\
  a_{21}^{(i)} \ \ & a_{22}^{(i)} \\
\end{array}%
\right)$, $\varepsilon_{x|y}$ and $\varepsilon_{y|x}$ are the
error terms, and we have assumed $m_1=m_2=m$ for simplicity. If
$m_1 \neq m_2$, some diagonal terms of $\mbf A_i$ would turn out
to be zero.

We note that Eq. (\ref{linear}) is just another form of Eq.
(\ref{granger}) for non-unit time step.  Therefore, in Eq.
(\ref{linear}), $\varepsilon_{x|y}$(or $\varepsilon_{y|x}$)
actually gives the prediction error of $x$ (or $y$) after
incorporating $y$ (or $x$). To proceed further, we also
reconstruct each series independently around the $x$ and $y$ parts
of the same reference point using linear regression approximations
to obtain

\parbox{10cm}{\beq x(t+\tau) & = & \sum_{j=1}^{m_1} \alpha_j
x[t+(j-1)\tau]+\varepsilon_x,\nonumber \\
y(t+\tau) & = & \sum_{j=1}^{m_2} \beta_j
y[t+(j-1)\tau]+\varepsilon_y. \nonumber \eeq} \hfill
\parbox{1cm}{\beq  \eeq}

We can now apply the ideas from Granger causality to these local
linear systems. Thus, if the ratio of the errors $\displaystyle
\frac{var(\varepsilon_{x|y})}{var(\varepsilon_x)}$ (or
$\displaystyle \frac{var(\varepsilon_{y|x})}{var(\varepsilon_y)}$)
is less than $1$, it implies $y$ (or $x$) has causal influence on
$x$ (or $y$). So far, this procedure only involves data in one
local neighborhood around a given reference point. Clearly, for
nonlinear systems the coefficient matrix in the linear
approximation is a function of the local neighborhood. We repeat
the process above for a set of chosen neighborhoods scattered over
the entire attractor and average the error ratios from all the
neighborhoods to obtain the Extended Granger Causality. See below
for a more formal formulation.

The idea above is actually very similar to other ideas of
detecting directional interdependence based on nonlinear mutual
prediction
\cite{arnhold,quian1,quian2,quyen,break1,break2,schiff,wies}. The
difference is that the previous work used the average or weighted
average value of the images of the points in a given neighbor as
the basis for prediction. Thus they suffer from the lacuna that
points in the neighborhood of the reference point have no explicit
time relations to the reference point itself.  On the other hand,
we employ a linear model which preserves explicit time relations
and are therefore able to derive unambiguous causality
relationships.

Summarizing, we propose the following four-step procedure to
evaluate causal relations between two nonlinear time series:
\begin{enumerate}
\item  Reconstruct the attractor using the delay coordinate
embedding technique [cf. Eq. (\ref{vector})]. \item  Fit an
autoregressive model for all the points in the neighborhood
$\Theta$ of a reference point $\mbf{z}_0$ in the reconstructed
space $\mbf R^{m_1+m_2}$, where $\Theta=\{\mbf{z}:
\left|\mbf{z}-\mbf{z}_0\right|\leq \delta\}$. \item  Perform the
reconstruction and fitting process on the individual $x$ and $y$
time series in the same neighborhood and compute the error ratio
defined earlier. Average the error ratio over a number of local
neighborhoods in order to sample the full attractor adequately.
Compute the Extended Granger Causality Index (EGCI) defined as
$\Delta_{y \rightarrow x}= \langle 1-\displaystyle
\frac{var(\varepsilon_{x|y})}{var(\varepsilon_x)} \rangle$, where
$\langle \cdot \rangle$ stands for averaging over the neighborhood
sampling the entire attractor. \item Compute EGCI as a function of
the neighborhood size $\delta$. For linear systems this index will
stay roughly the same as $\delta$ becomes smaller. For nonlinear
systems this index, in the small $\delta$ limit, reveals the true
nonlinear causal relation which may or may not be captured at the
full attractor level (i.e. taking $\delta$ to be the size of the
whole attractor).
\end{enumerate}

To reconstruct the attractor, the embedding dimension and time
delay have to be determined. Usually the embedding dimension is
determined by the false nearest neighbor technique \cite{kennel}
and the time delay is obtained as the first minimum of the mutual
information function \cite{fraser}. If the reconstructed attractor
is a fixed point with some noise, then a criterion such as AIC
\cite{akaike} for linear stochastic processes can be used to
determine the order of the model. A difficult issue for the
present work is to determine the optimal neighborhood size
$\delta$. The number of points in the neighborhood should be large
enough to establish good statistics. On the other hand, the size
of the neighborhood should be small enough so that linearization
is valid. In step (4) above we seek a compromise by examining the
EGCI as a function of $\delta$ in the attempt to get to the true
nonlinear effect when $\delta$ becomes small enough. We refer to
this step as a zooming-in procedure. For sufficiently large
dataset, the usual rule is that, the smaller the neighborhood
size, the better the nonlinear prediction achieved
\cite{break1,break2,wies}.

{\em Conditional Extended Granger Causality:} The above analysis
for two time series can be extended to more than two time series
by analyzing them pairwise. However, pairwise analysis of more
than two time series cannot detect indirect causal influences, an
issue that has been addressed in linear time series analysis
\cite{ding}. For example, consider three time series A, B and C.
Two possible causal relations among them are shown in Figures 1(a)
and (b).  In Figure 1(a), the causal influence or driving from A
to C is indirect and mediated by B. In Figure 1(b), both direct
and indirect influences exist. Pairwise analysis would show an
arrow from A to C and thus cannot separate these two cases. We
propose the following procedure for the case of three time series
which, as we show in the next section, is able to reveal the true
patterns of causal interactions.

Suppose $s_A(t),s_B(t),s_C(t)$ are the given three time series, we reconstruct
vector $\mbf z$ in whole space as\cite{pecora}:
\be \label{mvector}
\mbf{z}(t)=(\mbf s_A(t)^T,\mbf s_B(t)^T,\mbf s_C(t)^T)^T,
\ee
where $\mbf
s_A(t)=(s_A(t),s_A(t-\tau),\cdots,s_A(t-(m_1-1)\tau))^T,\ \ \mbf
s_B(t)=(s_B(t),s_B(t-\tau), \cdots,s_B(t-(m_2-1)\tau))^T,\ \ \mbf
s_C(t)^T=(s_C(t),s_C(t-\tau), \cdots,s_C(t-(m_3-1)\tau))^T.$
Then the vector autoregression obtained from a local linear
approximation is given by

\parbox{10cm}{\beq s_A(t+\tau)= \sum_{i=1}^{m_1} a_{11}^{(i)}
s_A[t-(i-1)\tau]+\sum_{i=1}^{m_2} a_{12}^{(i)} s_B[t-(i-1)\tau]
\nonumber \\ +\sum_{i=1}^{m_3} a_{13}^{(i)}
s_C[t-(i-1)\tau] + \varepsilon_{A|BC},\nonumber \\
s_B(t+\tau)= \sum_{i=1}^{m_1} a_{21}^{(i)}
s_A[t-(i-1)\tau]+\sum_{i=1}^{m_2} a_{22}^{(i)} s_B[t-(i-1)\tau] \nonumber \\
+\sum_{i=1}^{m_3} a_{23}^{(i)} s_C[t-(i-1)\tau] +
\varepsilon_{B|AC},\nonumber \\
s_C(t+\tau)= \sum_{i=1}^{m_1} a_{31}^{(i)}
s_A[t-(i-1)\tau]+\sum_{i=1}^{m_2} a_{32}^{(i)} s_B[t-(i-1)\tau]
\nonumber \\ +\sum_{i=1}^{m_3} a_{33}^{(i)} s_C[t-(i-1)\tau] +
\varepsilon_{C|AB}.\nonumber \eeq} \hfill
\parbox{1cm}{\beq  \eeq}

The term $\varepsilon_{C|AB}$ is the prediction error of the
series $\mbf s_C$ after incorporating both $\mbf s_A$ and $\mbf
s_B$. If this prediction is no better than the prediction obtained
by incorporating only the series $\mbf s_B$, it means $\mbf s_A$
has no direct causal influence on $\mbf s_C$. Therefore we can
define a Conditional Extended Granger Causality Index (CEGCI) as
$\Delta_{A \rightarrow C|B}= \langle~1- \displaystyle
\frac{var(\varepsilon_{C|AB})}{var(\varepsilon_{C|B})} \rangle$.
This can be used to distinguish between direct and indirect causal
relations. Conditional causality indices between other time series
pairs can be similarly defined.

It is worth mentioning that for more than three time series, if
the causality between any two time series is indirect, then taking
one additional series or more than one additional series in the
causality chain as the conditional one(s) will not make the
results any different. Therefore analysis of three time series is
sufficient to detect the intrinsic causal influences in any
multiple time series system.

\section{Numerical Simulations and Discussion}
In order to make the whole discussion concrete, we study some
examples. The number of reference points around the attractor is
100 for all the examples.

{\em Example 1:}  Let's consider two time series generated from
unidirectionally coupled 2D maps. Two different coupling schemes,
one linear and one nonlinear, are examined. The system with linear
coupling is written as

\parbox{10cm}{\beq \label{lin}
x(n) & = & 3.4x(n-1)(1-x^2(n-1))e^{-x^2(n-1)}+0.8x(n-2),\nonumber \\
y(n) & = & 3.4y(n-1)(1-y^2(n-1))e^{-y^2(n-1)}+0.5y(n-2)+c x(n-2),
\nonumber \eeq} \hfill
\parbox{1cm}{\beq  \eeq}

and the system with nonlinear coupling is

\parbox{10cm}{\beq \label{nonlin} x(n) & = &
3.4x(n-1)(1-x^2(n-1))e^{-x^2(n-1)}+0.8x(n-2), \nonumber \\
y(n) & = & 3.4y(n-1)(1-y^2(n-1))e^{-y^2(n-1)}+0.5y(n-2)+c
x^2(n-2).  \nonumber \eeq} \hfill
\parbox{1cm}{\beq  \eeq}

It is obvious that $y$ is driven by $x$ in both systems. In order
to make the simulations realistic, some system noise and
measurement noise are added to the time series. The attractor
reconstructed from the $x$ time series is given in Figure 2(a).
Figures 2(b) and 2(c) give the reconstructed attractors from the
$y$ time series driven linearly and nonlinearly by $x$ with the
coupling strength $c=0.5$.

Both these cases are analyzed using our procedure in the previous
section with $m=2$ and $\tau=1$. We obtain the EGCI as a function
of the neighborhood size $\delta$ in Figure 3. For both linear
driving [Fig. 3(a)] and nonlinear driving [Fig. 3(b)], $\Delta_{y
\rightarrow x}$ (shown as the thicker curve) is seen to be zero.
Thus $x$ is not influenced by $y$ as expected from the
construction of our model. In Fig. 3(a), $\Delta_{x \rightarrow
y}$ is non-zero starting from large neighborhood sizes $\delta$
and increases as $\delta$ decreases. This implies that $x$ has a
causal influence on or drives $y$. Further, since $\Delta_{x
\rightarrow y}$ is non-zero even for large $\delta$ values, this
means that even a linear causality analysis would have detected
the correct causal relations in this case. On the other hand, in
Figure 3(b) (for nonlinear driving), $\Delta_{x \rightarrow y}$
becomes non-zero only when $\delta$ is sufficiently small. In this
case, a linear causality analysis would fail to detect the correct
pattern of driving. This example illustrates the importance of
nonlinear causality analysis in such cases.

{\em Example 2:}  We consider two time series generated by two
coupled two-dimensional ODEs where the fixed point in the origin
is stable:

\parbox{10cm}{\beq
\dot{x_1} & = & -0.25x_1+x_2-x_2^3,\nonumber \\
\dot{x_2} & = & x_1-x_2-x_1x_2,\nonumber \\
\dot{y_1} & = & -0.25y_1+y_2-y_2^3+c x_1^2, \nonumber \\
\dot{y_2} & = & y_1-y_2-y_1 y_2. \nonumber \eeq} \hfill
\parbox{1cm}{\beq  \eeq}

Adding some system noise and measurement noise and taking $x_1$
and $y_1$ as the acquired signals, we get two modified time series
$x$ and $y$. Clearly in this case $x$ series drives $y$ series.
Reconstructing the attractors from these two time series with
$\tau=2$, finding the neighborhood of every reference point and
fitting a second order AR model ($m=2$) in every neighborhood, we
obtain the Extended Granger Causality Index (EGCI) as a function
of the neighborhood size $\delta$ for different coupling strengths
as shown in Figure 4(a). We make three observations. First,
$\Delta_{y \rightarrow x}\approx 0$, whereas $\Delta_{x
\rightarrow y}$ is non-zero, clearly demonstrating that the $x$
series drives the $y$ series but not vice-versa. Second, the level
of EGCI is proportional to the coupling strength. Third, since the
processes here are linear, the EGCI is not a function of $\delta$.

{\em Example 3:} Next we look at an ODE system exhibiting chaotic
behaviors. The following two coupled R\"{o}ssler oscillators are
considered:

\parbox{10cm}{\beq \dot{x_1} & = &
-(y_1+z_1),\nonumber \\
\dot{y_1} & = & x_1+0.2 y_1,\nonumber \\
\dot{z_1} & = & 0.2+z_1(x_1-4.7). \nonumber \\
\dot{x_2} & = &
-(y_2+z_2)+c x_1,\nonumber \\
\dot{y_2} & = & x_2+0.2 y_2,\nonumber \\
\dot{z_2} & = & 0.2+z_2(x_2-4.7). \nonumber \eeq} \hfill
\parbox{1cm}{\beq  \eeq}

As done earlier, some system noise and measurement noise are added
to the two time series $x_1$ and $x_2$ to obtain $x$ and $y$ time
series. Reconstructing the attractors with $m=3$ and $\tau=2$ and
fitting linear models in every local neighborhood, we obtain EGCI
shown in Figure 4(b). It is seen that $x$ has a causal influence
on $y$ as expected. Besides, $\Delta_{x \rightarrow y}$ is
non-zero even for large neighborhood sizes. Thus, a linear
causality analysis would detect the correct causal relations in
this strongly nonlinear system.

{\em Example 4:}  To show how to distinguish the pattern of
interaction shown in Figure 1(b) from that shown in Figure 1(a),
let us consider three time series. Both linear systems and
nonlinear systems are considered.

For a linear stochastic system the following three coupled AR(1)
models are considered:

\parbox{10cm}{\beq x(n) & = &
0.2x(n-1)+\varepsilon_1,\nonumber \\
y(n) & = & 0.5y(n-1)+0.5x(n-1)+\varepsilon_2, \nonumber \\
z(n) & = & 0.4z(n-1)+0.3y(n-1)+c x(n-1)+\varepsilon_3. \nonumber
\eeq} \hfill
\parbox{1cm}{\beq  \eeq}

For chaotic time series, the following three coupled 1-d maps are
considered:

\parbox{10cm}{\beq x(n) & = &
3.4x(n-1)(1-x^2(n-1))e^{-x^2(n-1)},\nonumber \\
y(n) & = & 3.4y(n-1)(1-y^2(n-1))e^{-y^2(n-1)}+0.5x(n-1), \nonumber \\
z(n) & = & 3.4z(n-1)(1-z^2(n-1))e^{-z^2(n-1)}+0.3y(n-1)+c x(n-1).
\nonumber \eeq} \hfill
\parbox{1cm}{\beq  \eeq}

Here $x$, $y$ and $z$ correspond to A, B and C in Figure 1. In
addition, $c=0$ simulates the indirect causality case (Fig. 1(a))
and $c=0.5$ simulates the direct causality case (Fig. 1(b)).

Numerical results for the linear case are shown in Figure 5 and
results for the chaotic time series are shown in Figure 6. Figures
5(a) and 6(a) display results obtained using pairwise analysis.
Similar plots are obtained for both $c=0$ and $c=0.5$. As we can
see, based on just pairwise analysis, one would conclude that
Figure 1(b) is the pattern of interaction for both $c=0$ and
$c=0.5$. That is, the direct and indirect causal relationships
cannot be separated by pairwise analysis alone. Figures 5(b) and
6(b) give the results obtained by simultaneously analyzing all
three time series using conditional EGCI. In this case, for $c=0$
we obtain $\Delta_{x \rightarrow z|y} \approx 0$ (solid curve)
indicating that no direct causal relation exists between $x$ and
$z$. Thus the correct causality graph [Figure 1(a)] is obtained.
On the other hand, for $c=0.5$, $\Delta_{x \rightarrow z|y}$ is
non-zero (dotted curve) indicating direct causality between $x$
and $z$ and the causality graph shown in Figure 1(b) is obtained.

\section{Conclusions}

We have extended the Granger causality theory to nonlinear time
series by incorporating the embedding reconstruction technique for
multivariate time series. A four-step algorithm was presented and
used to analyze various linear and nonlinear coupled systems. The
following conclusions were found:
\begin{enumerate}
\item Linear Granger causality analysis may or may not work for
nonlinear time series. On the other hand, our method of applying
the Extended Granger Causality Index to nonlinear time series
always gives reliable results. \item When three or more time
series have to be analyzed, the Conditional Extended Granger
Causality Index proposed here can distinguish between direct and
indirect causal relationships between any two of the time series.
This is not possible using simple pairwise analysis.
\end{enumerate}
As with other methods for analyzing nonlinear time series, the amount of data
required for reliable analysis can be large. Possible improvements
along this direction are being studied.
\newpage

\section*{Acknowledgments}

The work was supported by ONR, NSF and NIMH. GR is also supported
by DRDO, India and the Homi Bhabha Fellowship.

\newpage
\section*{Figure Caption}

\begin{description}
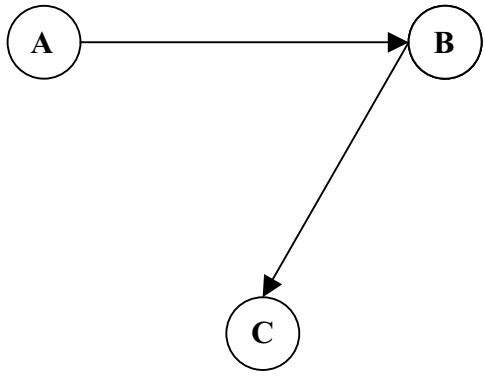
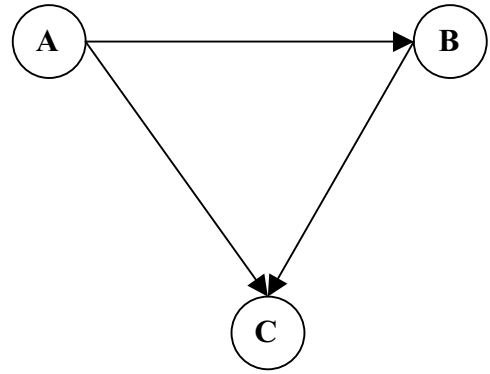
\item{\bf Figure 1:} Two patterns of causal interactions. (a) A
drives C by way of B and (b) There is a direct pathway from A to
C.

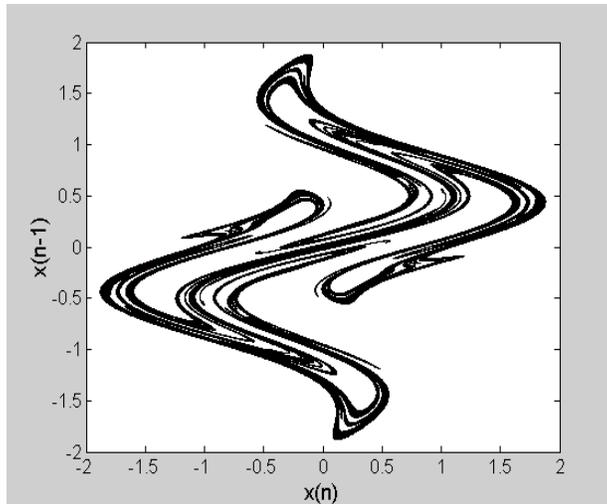
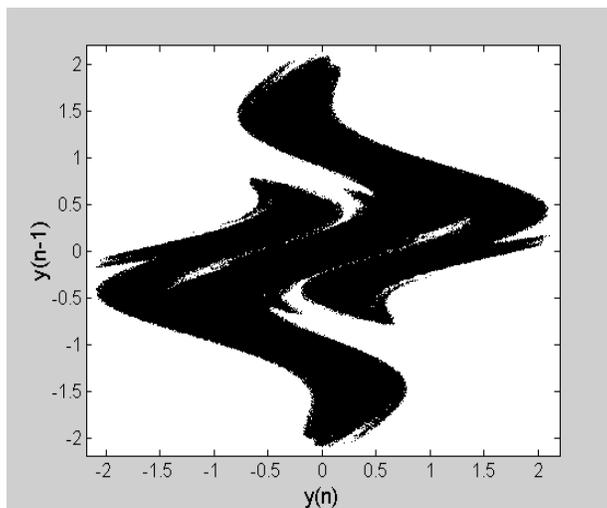
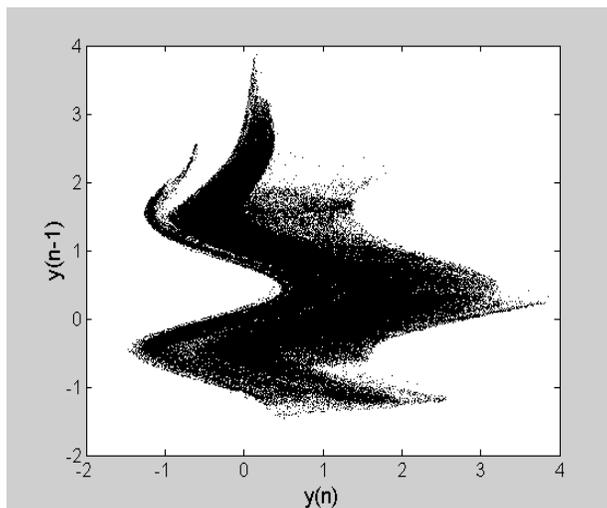
\item{\bf Figure 2:} Reconstructed attractors from time
series from Example 1.  (a) Driving attractor; (b) linearly
driven attractor; (c) nonlinearly driven attractor.

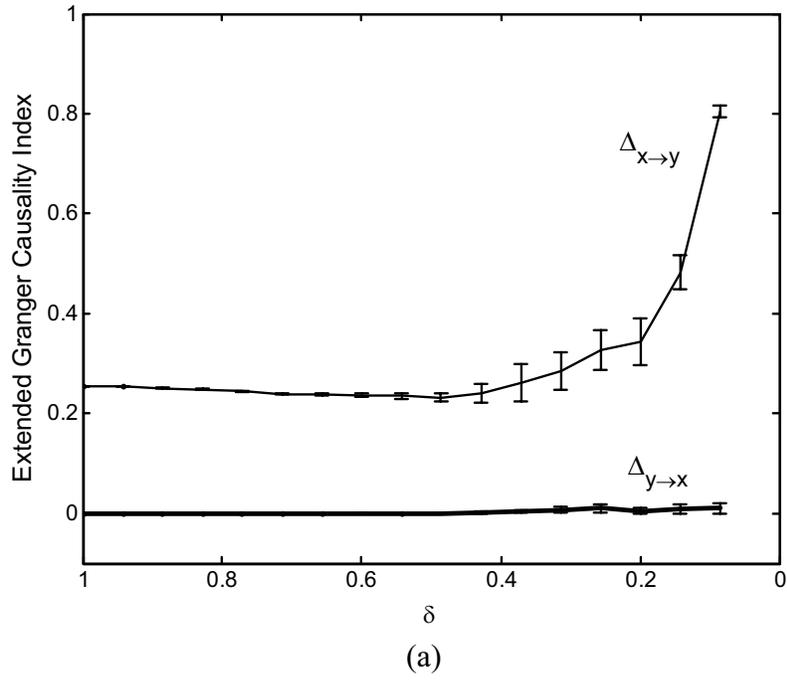
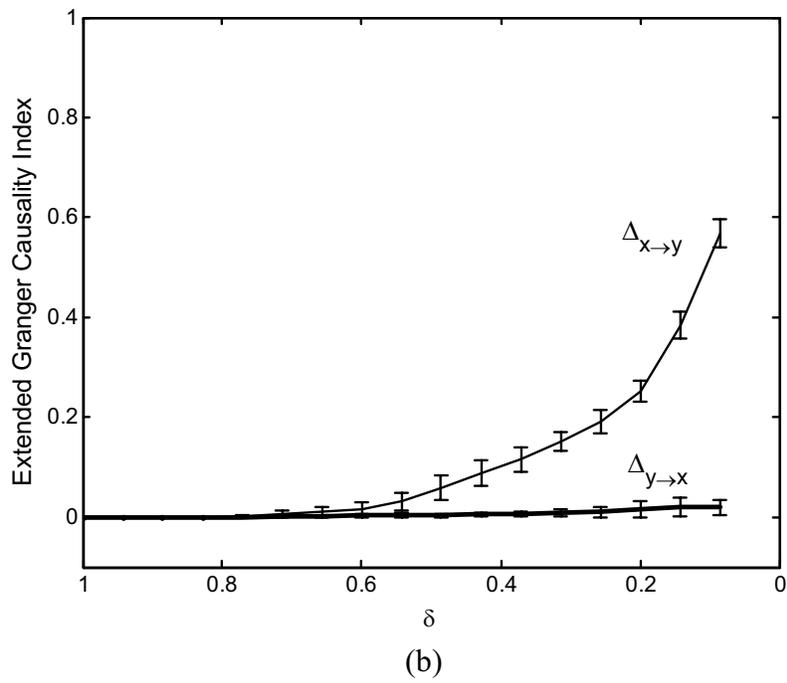
\item{\bf Figure 3:} Extended Granger Causality Index (EGCI) as a
function of the size $\delta$ of the neighborhood from Example 1.
(a) Linear driving case; (b) nonlinear driving case.

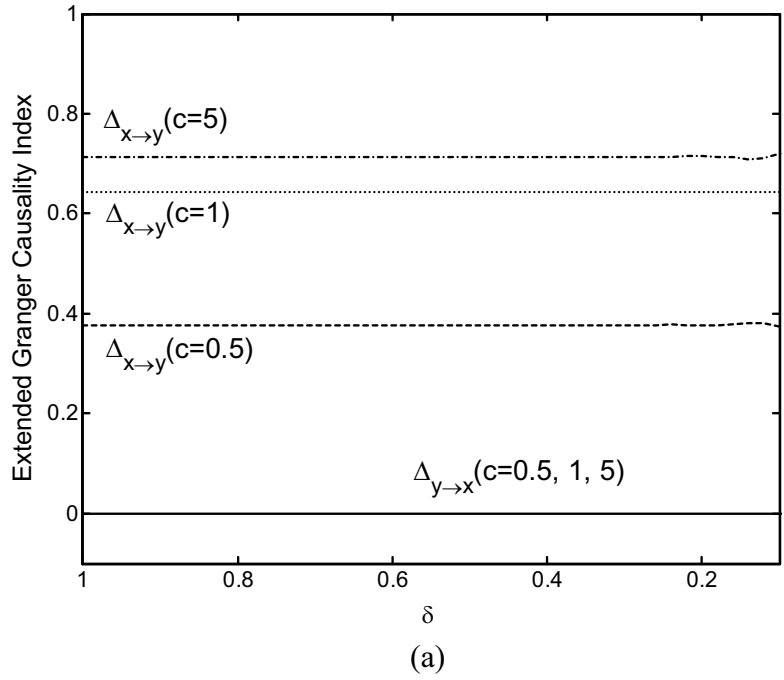
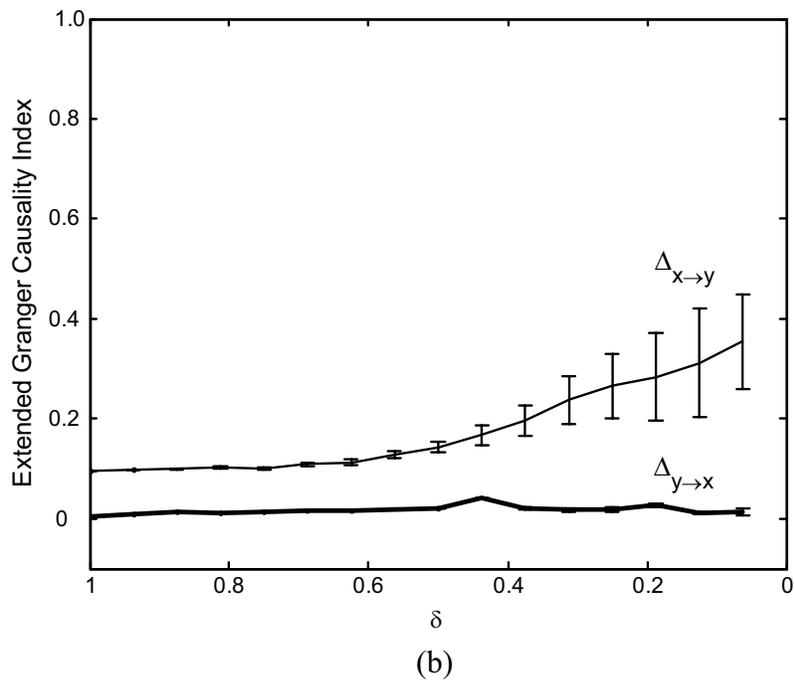
\item{\bf Figure 4:} EGCI between two time series from Examples 2
and 3. (a) ODEs with a stable fixed point and different coupling
strengths; (b) ODEs with chaotic behavior.

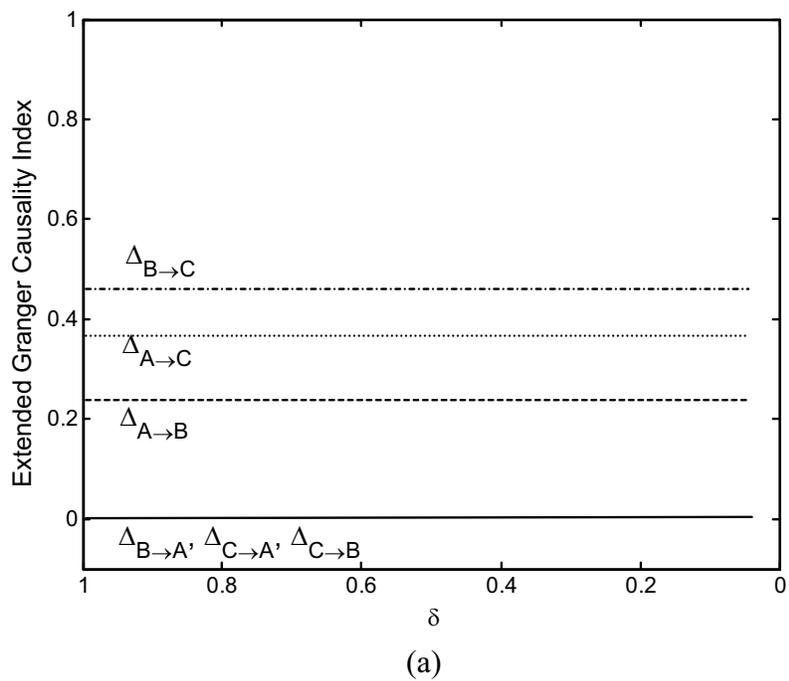
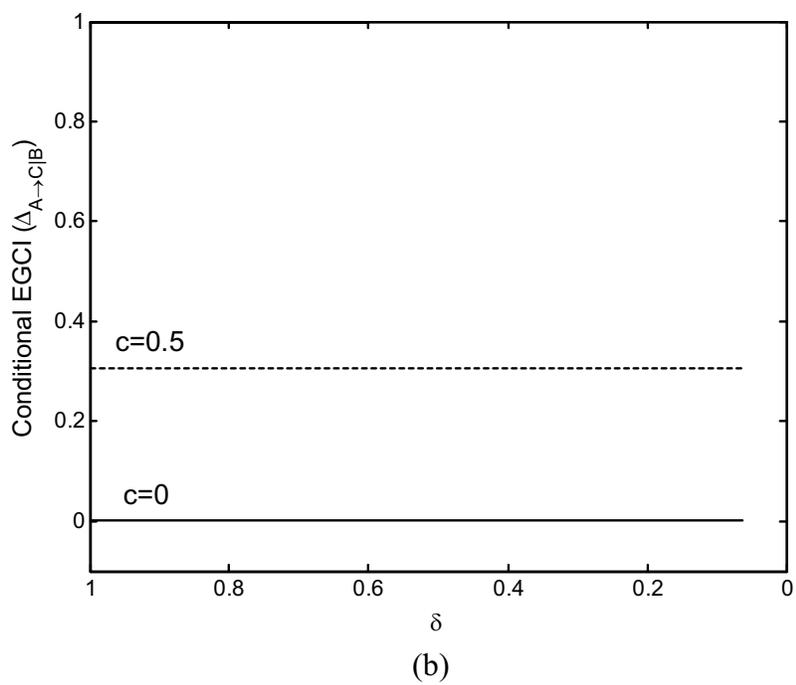
\item{\bf Figure 5:} Simulating two patterns of interactions in
Figure 1 with three coupled AR models (Example 4). (a) Pairwise
analysis results; (b) Conditional causality analysis separates the
two cases. Dotted line gives the Conditional Extended Granger
Causality Index (CEGCI) for $c=0.5$ and the solid line for $c=0$.

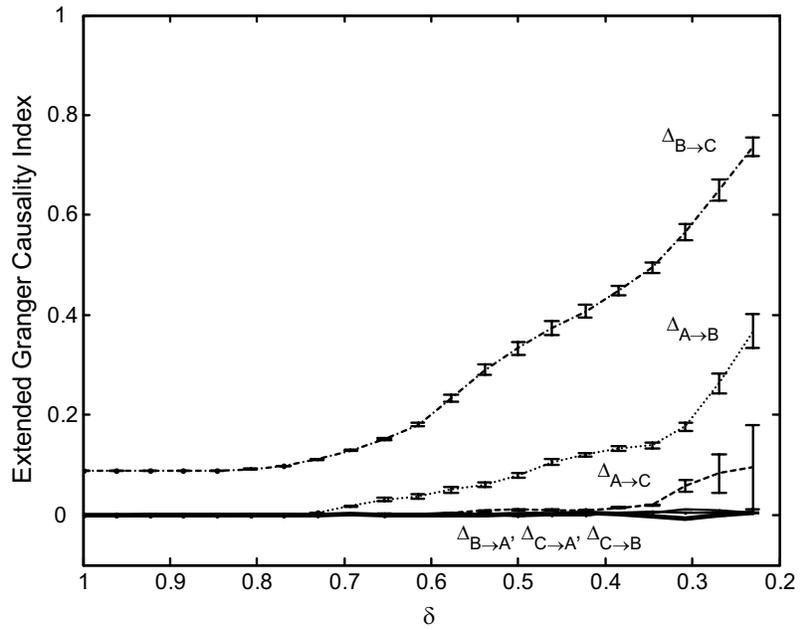
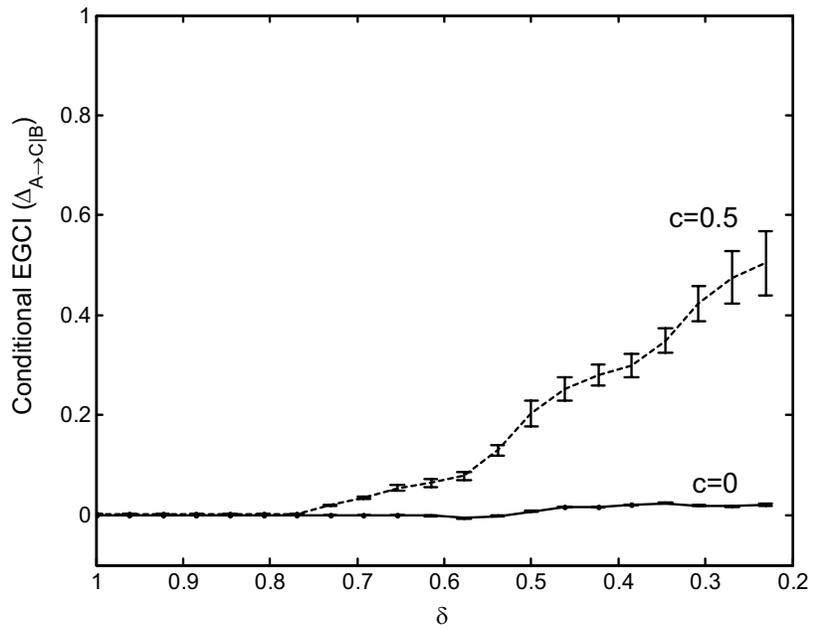
\item{\bf Figure 6:} Simulating two patterns of interactions in
Figure 1 with three coupled chaotic 1d-maps (Example 4). (a)
Pairwise analysis results, the two patterns of interaction are not
distinguished; (b) CEGCI analysis is able to distinguish between
the two different patterns. Dotted line gives the CEGCI for
$c=0.5$ and the solid line for $c=0$.

\end{description}

\end{document}